\documentclass[preprint,showpacs,preprintnumbers,amsmath,amssymb]{revtex4}

\usepackage{graphicx}
\usepackage{dcolumn}
\usepackage{bm}
\usepackage{epsfig}
\allowdisplaybreaks[4]

\begin{document}

\title{Bethe Ansatz for 1D  interacting anyons}
\author{M.T. Batchelor$^{\dagger}$,  X.-W. Guan$^{\dagger}$ and J.-S. He$^{\ddagger}$}
\affiliation{$^{\dagger}$Department of Theoretical Physics, 
Research School of Physical Sciences and Engineering, and\\
Mathematical Sciences Institute,
Australian National University, Canberra ACT 0200,  Australia\\
$^{\ddagger}$Department of Mathematics, University of Science and
Technology of China, Hefei 230026, Anhui, P. R. China}

\date{\today}

\begin{abstract}
\noindent
This article gives a pedagogic derivation of the Bethe Ansatz solution for 1D interacting
anyons. This includes a demonstration of the subtle role of the anyonic
phases in the Bethe Ansatz arising from the anyonic commutation relations.  
The thermodynamic Bethe Ansatz equations defining the temperature dependent
properties of the model are  also derived, from which some groundstate properties are 
obtained.

\end{abstract}



\maketitle
\section{Introduction}
\label{Intro}

A number of 1D models in quantum many-body physics have been solved by the Bethe Ansatz 
following Bethe's pioneering work on the exact solution of the 1D Heisenberg
magnetic spin chain in 1931 \cite{Bethe,Batchelor}.
During the past 75 years the Bethe Ansatz has been developed and applied to physical problems such as the
1D $\delta$-function interacting Bose \cite{LL,McQuire} and Fermi 
\cite{Yang-Gaudin} gases, the 1D Hubbard model \cite{Lieb-Wu} and 
2D vertex models \cite{Lieb-Wu2,Baxter} (many of these key papers are collected 
in Ref.~\cite{Mattis}).
The solution of such models contributed to the development of the 
celebrated Yang-Baxter equation, which gives
the consistency conditions for many-body scattering problems 
and plays a crucial role in quantum integrable systems
\cite{Gaudin-B,Korepin-B,Sutherland-B,Takahashi-B,Ma-B,Hubbard-B} and in
modern mathematical physics \cite{Jimbo,Drinfeld}.  
The related establishment
of the quantum inverse scattering method \cite{Korepin-B,QISM1,QISM2} 
provided widespread applications of the Yang-Baxter equation in 
low-dimensional quantum systems, such as the Kondo problem \cite{Kondo},
the Anderson model \cite{Anderson} and long range interaction systems \cite{Sierra,Zhou}.

A further pioneering application of the Bethe Ansatz was to the
1D $\delta$-function interacting Bose gas \cite{LL} which is
related to the quantum nonlinear Schr\"{o}dinger equation. 
This model provides an important realistic physical description of an
interacting 1D Bose gas.
It is also arguably one of the most simplest and pedagogic models 
solved in terms of the Bethe Ansatz. 
In general the applicability of the
Bethe Ansatz depends on the reducibility of the multi-particle scattering
matrix to the product of many two-particle scattering matrices. 
The starting point for the Bethe Ansatz approach in quantum many-body physics
is to reduce the eigenvalue problem of the field theoretic hamiltonian
into a quantum mechanical many-body problem.  
The wavefunction of the many-body hamiltonian inherits the statistical signature of 
the interacting particles, which leads to striking and subtle quantum many-body effects. 
For example, significantly different quantum effects between the 1D
$\delta$-function interacting Bose and Fermi gases are seen clearly
from the Bethe Ansatz solutions \cite{LL,Yang-Gaudin}.

On the other hand, anyons \cite{F-S,F-S2} may also exist in 
both two and one dimension, obeying fractional statistics. 
For a 2D electron gas in the fractional quantum Hall (FQH) regime, 
the quasi-particles are charged anyons \cite{cond1}.  
A more general description of quantum statistics
is provided by Haldane exclusion statistics \cite{Haldane}, which is a 
formulation of fractional statistics based on a generalized Pauli exclusion principle, 
now called generalized exclusion statistics \cite{Wu,Wilczek}. 
In 1D, a wavefunction with anyonic symmetry may also occur \cite{Kundu}. 
Anyons in 1D acquire a multi-step function-like phase when two identical 
particles exchange their positions in the scattering process. 
This topological phase results in rich quantum effects in the 1D interacting model of anyons
\cite{Zhu,BGO,BG,Girardeau,Roaul}.  

In this paper, celebrating the 75th anniversary of the Bethe Ansatz, 
we give a detailed and pedagogic derivation of the Bethe Ansatz solution for  
1D $\delta$-function interacting anyons, discussed for the first time by Kundu \cite{Kundu}. 
The ground state properties derived from the thermodynamic
Bethe Ansatz are also presented.

This paper is set out as follows. In section \ref{model}, we introduce
the 1D interacting anyon model and show the reducibility of the
field-theoretic Hamiltonian to a quantum mechanics many-body problem.
The Bethe Ansatz solution is derived in section \ref{sec:BA}.  In section
\ref{sec:TBA}, we study the thermodynamic Bethe Ansatz equations (TBA).  Section
\ref{conclusion} is devoted to concluding remarks.

\section{The model}
\label{model}

We consider $N$ anyons with $\delta$-function interaction in 1D with hamiltonian \cite{Kundu}
\begin{eqnarray}
H &=&\frac{\hbar^2}{2m}\int_0^Ldx\,\partial \Psi ^{\dagger}(x) \partial
  \Psi(x) +\frac12 \, g_{\rm 1D} \int_0^Ldx\,\Psi^{\dagger}(x)\Psi^{\dagger}(x)\Psi(x) \Psi(x)
\label{Ham1}
\end{eqnarray}
under periodic boundary conditions, with $x$ a coordinate in length $L$. 
Here $m$ denotes the atomic mass.
Hereafter we set $\hbar =2m=1$ for convenience.  
The coupling constant $g_{\rm 1D}$ is determined by $g_{\rm 1D}={\hbar^2c}/{m}$ where, 
at least for the Bose gas, the coupling strength $c$ is tuned through an effective $1$D scattering
length $a_{1D}$ via confinement in experiments.
We also use a dimensionless coupling constant $\gamma =c/n$ to characterize different 
physical regimes of the anyon gas, where $n=N/L$ is the linear density.  
The operators $\Psi^{\dagger}(x)$ and $\Psi(x) $ are the creation and annihilation
operators (or operator valued particle density) at point $x$
satisfying the anyonic (equal-time) commutation relations 
\begin{eqnarray}
\Psi (x_1)\Psi ^{\dagger}(x_2)&=&{\mathrm e}^{-\mathrm{i}\kappa
  w(x_1,x_2)}\Psi ^{\dagger}(x_2)\Psi(x_1)+\delta(x_1-x_2)\nonumber\\
\Psi (x_1)\Psi (x_2)&=&{\mathrm e}^{\mathrm{i}\kappa
  w(x_1,x_2)}\Psi (x_2)\Psi(x_1), \nonumber\\
\Psi ^{\dagger}(x_1)\Psi ^{\dagger}(x_2)&=&{\mathrm e}^{\mathrm{i}\kappa
  w(x_1,x_2)}\Psi ^{\dagger}(x_2)\Psi^{\dagger}(x_1).\label{comm}
\end{eqnarray}
Here the multi-step function $w(x_1,x_2)=-w(x_2,x_1)=1$ for order $x_1>x_2$,
with $w(x,x)=0$.  
The anyonic phase $w(x,x)=0$ for two colliding
particles has bosonic signature at the  point $x_1=x_2$. 
We note that in imposing the restriction 
${\mathrm e}^{\mathrm{i} \kappa w(x,x)}={\mathrm e}^{\mathrm{i}\pi }$ 
on the anyonic phase the above commutation relations give hard-core
relations which have been constructed based on lattice
integrable models \cite{LIM} and also suggested in the anyon-Fermi
mapping for the anyonic Tonks-Girardeau gas \cite{Girardeau}. 
We shall see that this phase is not suitable for the continuous model of anyons. 
In contrast to the $1$D interacting Bose gas \cite{LL}, hamiltonian (\ref{Ham1}) exhibits both 
anyonic statistical and dynamical interactions, which can map to a $1$D interacting 
Bose gas with multi-$\delta$-function and momentum-dependent interactions \cite{Kundu}. 
The more general anyonic statistical and dynamical interactions result in a richer range of 
quantum effects than those of the 1D interacting Bose gas.

We first consider the corresponding equation of motion 
$-\mathrm{i}\partial _t \Psi (x,t) =\left[H, \Psi (x,t) \right]$ via the
time-dependent quantum fields, i.e., the nonlinear Schr\"{o}dinger equation
\begin{equation}
\mathrm{i} \partial _t \Psi (x,t)=-\partial _x^2\Psi
(x,t)+g_{\rm 1D}\Psi^+ (x,t) \Psi^2 (x,t). \label{NSE}
\end{equation}
In order to appreciate  the subtlety of the anyonic phase, the 
nonlinear Schr\"{o}dinger equation can be calculated as 
\begin{eqnarray}
& & \left[\int_0^Ldx_1 \left( \partial_{x_1} \Psi ^{\dagger}(x_1)
\partial_{x_1} \Psi(x_1)+\frac{1}{2} g_{\rm
1D}(\Psi^{\dagger}(x_1))^2(\Psi(x_1))^2\right),
\Psi(x_2) \right]\nonumber\\
=& &\left[\int_0^Ldx_1 \left( -\Psi^{\dagger}(x_1) \partial^2_{x_1} \Psi(x_1)+
\frac{1}{2} g_{\rm 1D}{\Psi^{\dagger}}^2(x_1)\Psi^2(x_1) \right),
\Psi(x_2) \right]\nonumber \\
= & &
-\int_0^Ldx_1 \left(
\mathrm{e} ^{\mathrm{i}\kappa w(x_1,x_2)}\Psi ^{\dagger}(x_1)\Psi(x_2) 
-\mathrm{e}^{-\mathrm{i}\kappa
w(x_2,x_1)}\Psi^{\dagger}(x_1)\Psi(x_2)-\delta(x_2-x_1)\right)\partial^2_{x_1}\Psi(x_1)\nonumber\\
& &
+ \, \frac12 g_{\rm 1D}\int_0^Ldx_1 \left[  {\Psi^{\dagger}}^2(x_1)\Psi^2(x_1)\Psi(x_2)-
\left( \mathrm{e}^{-2\mathrm{i}\kappa w(x_2,x_1)}
{\Psi^{\dagger}}^2(x_1) \Psi(x_2)\right. \right. \nonumber\\
& &  + \left. \left.
\mathrm{e}^{-\mathrm{i}\kappa w(x_2,x_1)}\delta(x_2-x_1)
\Psi^{\dagger}(x_1)+ \mathrm{e}^{-2\mathrm{i}\kappa
w(x_2,x_1)}\delta(x_2-x_1) \Psi^{\dagger}(x_1) \right) \Psi^2(x_1) \right]\nonumber\\
=& & \partial_{x_2}^2\Psi(x_2)-g_{1D}\Psi^{\dagger}(x_2)\Psi^2(x_2).\label{motion}
\end{eqnarray}
In the above equation the time $t$ has been omitted for simplicity. 
The properties $w(x,x)=0$ and $w(x_1,x_2)=-w(x_2,x_1)$ have been used.
%
%
We stress that the phases induced from exchanging operator positions in the terms 
$\Psi(x_1)\Psi(x_2)$ and $\Psi(x_2)\Psi^{\dagger} (x_1)$ are some fixed constants, 
depending on the relation between $x_1$ and $x_2$. 
The derivative operator $\partial _{x_1}$ then only acts on the field operators.  
Furthermore, if we choose ${\mathrm e}^{\mathrm{i} \kappa w(x,x)}={\mathrm e}^{\mathrm{i}\pi }$, the last term in
the above equation does not exhist due to the cancellation of the
terms $\mathrm{e}^{-\mathrm{i}\kappa w(x_2,x_1)}\delta(x_2-x_1)
\Psi^{\dagger}(x_1)$ and $\mathrm{e}^{-2\mathrm{i}\kappa
w(x_2,x_1)}\delta(x_2-x_1) \Psi^{\dagger}(x_1)$. 
This means that there is no $\delta$-function interaction in the
quantum mechanical many-body hamiltonian without internal spin degrees
of freedom due to the hard-core behaviour.
For this reason we take the choice of phase $w(x,x)=0$ in order to
ensure  the commutation relations (\ref{comm}). Under this requirement
we can obtain  the form of the nonlinear Schr\"{o}dinger equation (\ref{NSE}).

Following the way suggested for the 1D Bose gas \cite{Korepin-B}, we  define a Fock vacuum
state $\Psi(x) \! \mid \! 0\rangle=0$. 
Thus the number operator ${\bf  N} $ and momentum operator ${\bf P}$ are
\begin{eqnarray}
{\bf N}& =&\int_0^Ldx \Psi^{\dagger}(x)\Psi(x), \\
{\bf P}& =& \mathrm{i}\int_0^Ldx \left[
\partial_x \Psi^{\dagger}(x)\right] \Psi(x). \label{P}
\end{eqnarray}
To properly denote the anyonic phases in the eigenfunctions $\mid \! \Phi \rangle$ 
of hamiltonian (\ref{Ham1}), we assign particle coordinates $x_i$ in the order
$x_1 < x_2 < \cdots < x_N$.  
Based on this assigned order the multi-step function $w(x_i,x_j)$ can 
be used to count phase alternation when two particles are 
interchanged, i.e., $w(x_i,x_j)=-w(x_j,x_i)$.  
The $N$-particle eigenstate is written as
\begin{equation}
\mid \! \Phi \rangle=\int_0^L dx^N {\mathrm e}^{-\mathrm{i}\frac{\kappa N}{2}}\chi
(x_1\ldots x_N)\Psi ^{\dagger}(x_1)\ldots \Psi^{\dagger}(x_N) \! \mid \! 0\rangle
\label{state}
\end{equation}
where the Bethe Ansatz wavefunction is of the form  
\begin{eqnarray}
\chi(x_1\ldots
x_N)&=&{\mathrm e}^{-\frac{\mathrm{i}\kappa}{2}\sum_{x_i<x_j}^Nw(x_i,x_j)}\sum_PA(k_{P1}\cdots
k_{PN}) {\mathrm e}^{\mathrm{i}(k_{P1}x_1+\ldots +k_{PN}x_N)}.
\label{wave}
\end{eqnarray}
Here the sum extends over all $N!$ permutations $P$. 
The order in which the particles are created in the $N$-particle eigenstate (\ref{state}) 
incurs the phase factor in the wave function (\ref{wave}). 
Integration involves changes in the order of 
creating particles due to the permutation of coordinates.  
We can denote the permutation $P$ as 
$$P= \left(\begin{array}{llll} 1&2&\ldots &N \\ P1&P2&\ldots &PN \end{array}\right) $$
which transforms the $m$-th object into the $Pm$-th position for
$1\leq m \leq N$ \cite{Ma-B}. 
In general, for two different permutations we have 
\begin{eqnarray}
Q=\left(\begin{array}{llll} 1&2&\ldots &N \\ Q1&Q2&\ldots &QN \end{array}\right)=\left(\begin{array}{llll} P1&P2&\ldots &PN \\ Q_{P1}&Q_{P2}&\ldots &Q_{PN} \end{array}\right).\nonumber
\end{eqnarray}
Therefore
\begin{eqnarray}
QP=\left(\begin{array}{llll} 1&2&\ldots &N \\ Q_{P1}&Q_{P2}&\ldots
  &Q_{PN} \end{array}\right),\,\,\, QP^{-1}=\left(\begin{array}{llll} P1&P2&\ldots &PN \\ Q_{1}&Q_{2}&\ldots
  &Q_{N} \end{array}\right).\nonumber
\end{eqnarray}

In order to understand the above expressions, we give the explicit
form of the eigenstate ({\ref{state}) for $N=3$ in Appendix
\ref{APPa}.
The coefficients $A(k_{P1}\ldots k_{PN})$ are obtained explicitly via the
Bethe Ansatz approach in the next section. 
By comparing eqns (\ref{w1}) and (\ref{w4}) we see clearly that the phase factors 
in the multi-valued wavefunction (\ref{wave}) are diminished by those from permutations
of the particles in the eigenstate $\mid \!\! \Phi \rangle$ such that
the integrand in (\ref{state}) is single valued and fully symmetric.
We extracted a global phase factor ${\mathrm e}^{-\mathrm{i}{\kappa N}/{2}}$ 
in order to symmetrize the anyonic phase factor in the wave
function (\ref{wave}) so that it has $\kappa \to \kappa +4\pi $ symmetry. 
The eigenstate still has $\kappa \to \kappa +2\pi $ symmetry. 
Moreover, it is easily seen that the wave function satisfies the anyonic symmetry \cite{Kundu}
\begin{equation}
\chi(\ldots x_i\ldots x_j \ldots )={\mathrm e}^{-\mathrm{i} \kappa
\left(\sum_{k=i+1}^jw(x_i,x_k)-\sum_{k=i+1}^{j-1}w(x_j,x_k) \right) }
\chi(\ldots x_j\ldots x_i \ldots ).\label{A-S}
\end{equation}

Acting on the eigenstate (\ref{state}) with the operator ${\bf P}$ defined in (\ref{P}) gives
\begin{eqnarray}
{\bf P}\mid \! \Phi \rangle&=&\mathrm{i}\int_0^L dy \int_0^L dx^N {\mathrm
  e}^{-\mathrm{i}\frac{\kappa N}{2}}\chi(x_1\ldots
  x_N)\left[\partial_y \Psi^+(y)\right] \nonumber\\
 & &\times \sum
  _{k=1}^N\mathrm{e}^{-\mathrm{i}\kappa\sum_{i=1}^{k-1} w(y-x_i)}\Psi
  ^{\dagger}(x_1)\ldots\Psi^{\dagger}(x_{k-1})\delta(y-x_k) \Psi
  ^{\dagger}(x_{k+1})\ldots \Psi ^{\dagger}(x_N) \! \mid \!
  0\rangle\nonumber \\
 &= &\mathrm{i}\int_0^L dx^N {\mathrm
  e}^{-\mathrm{i}\frac{\kappa N}{2}}\chi(x_1\ldots x_N)\nonumber\\ &
  &\times \sum _{k=1}^N\Psi
  ^{\dagger}(x_1)\ldots\Psi^{\dagger}(x_{k-1})\partial_{x_k}\Psi
  ^{\dagger}(x_{k}) \Psi^{\dagger}(x_{k+1})\ldots \Psi ^{\dagger}(x_N)
  \! \mid \! 0\rangle\nonumber\\
 &=&\int_0^L dx^N {\mathrm
  e}^{-\mathrm{i}\frac{\kappa N}{2}}\left\{-\mathrm{i}\sum
  _{k=1}^N\partial _{x_k} \chi(x_1\ldots x_N)\right\} \Psi
  ^{\dagger}(x_1)\ldots \Psi ^{\dagger}(x_N) \! \mid \! 0\rangle
\end{eqnarray}
It therefore follows that we can obtain the quantum mechanical momentum operator as
\begin{eqnarray}
{\bf P}&=&-\mathrm{i}\sum_{k=1}^N\partial _{x_k}.
\end{eqnarray}

Similarly, by acting on the eigenstate (\ref{state}) with
the hamiltonian (\ref{Ham1}) the eigenvalue problem for hamiltonian (\ref{Ham1}), namely
\begin{eqnarray}
H\mid \! \Phi \rangle=\int_0^L dx^N {\mathrm
  e}^{-\mathrm{i}\frac{\kappa N}{2}}
H_N\chi(x_1\ldots x_N)\Psi ^{\dagger}(x_1)\ldots \Psi
^{\dagger}(x_N) \! \mid \! 0\rangle,
\end{eqnarray}
can be reduced to solving the quantum mechanical problem
\begin{equation}
H_N\chi (x_1\ldots x_N)=E\chi (x_1\ldots x_N),\label{Q-E}
\end{equation}
where
\begin{equation}
H_N=-\frac{\hbar ^2}{2m}\sum_{i = 1}^{N}\frac{\partial
^2}{\partial x_i^2}+\,g_{\rm 1D}\sum_{1\leq i<j\leq N} \delta
(x_i-x_j). \label{Ham2}
\end{equation}
The quantum mechanical hamiltonian (\ref{Ham2}) describes the $1$D
$\delta$-function interacting quantum gas of $N$ anyons confined in a periodic length $L$. 
The details of the above calculation are lengthy but straightforward. 
Without loss of generality, we give the details of the proof for the case $N=3$ in Appendix \ref{APP1}. 
Again, the subtlety of the anyonic phases is such that they just match the applicability of the Bethe
Ansatz -- from the eigenvalue problem of the field theoretic hamiltonian
(\ref{Ham1}) to the quantum many-body problem (\ref{Ham2}). 
The next task is to determine the wavefunction by means of the Bethe Ansatz.


\section{Bethe Ansatz solution}
\label{sec:BA}

In the preceding section, we showed the equivalence between the quantum field theoretic
and quantum many-body problems. 
The Bethe Ansatz wavefunction (\ref{wave}) was written for the fundamental
domain $0\le x_1<x_2< \ldots <x_N \le L $ \cite{Kundu,BGO}. 
The wavefunction in other domains is an extension of the wavefunction (\ref{wave}) 
via the anyonic symmetry (\ref{A-S}). 
The $\delta$-function potential causes a jump in the derivative of the
wavefunction in the eigenvalue equation (\ref{Q-E}).
Changing the coordinates to centre of mass coordinates 
$X=({x_j+x_k})/{2}$ and $Y=x_j-x_k$ leads to the eigenvalue equation (\ref{Q-E}) 
in the form
\begin{equation}
\left\{ \left(-\frac{\partial^2}{\partial x_1^2}\cdots
-\frac{1}{2}\frac{\partial^2}{\partial
  X^2}-2\frac{\partial^2}{\partial Y^2}\cdots -\frac{\partial
  ^2}{\partial x_N^2}\right) +2c\delta(Y)-E \right\}\chi(\cdots x_i\cdots x_j \cdots )=0.
\end{equation}
Integrating both sides of this  equation with respect to $Y$ from
$-\epsilon$ to $+\epsilon$ and taking the limit ${\epsilon \to 0}$ 
gives the discontinuity condition
\begin{eqnarray}
& &\left( \partial _{x_{j}}-\partial _{x_i}
\right)\chi(x_1,\ldots,
x_i,x_{j},\ldots,x_N)|_{x_j=x_i+\epsilon} -\left( \partial _{x_j}-\partial _{x_i}
\right)\chi(x_1,\ldots,
x_j,x_{i},\ldots,x_N)|_{x_j=x_i-\epsilon}\nonumber\\
& &
=2c\chi(x_1,\ldots, x_i,x_{j},\ldots,x_N)|_{x_i=x_j}.\label{jump}
\end{eqnarray}
on the derivative of the  wavefunction.
The two particle scattering relation can then be worked out directly
from the wavefunction $\chi(x_1,\ldots ,x_N)$.  
This model does not have  internal degrees of freedom.  Therefore, it is not
necessary to introduce extra coordinate indices to distinguish the
coefficients in the wave function for different domains.  
We may write the wavefunction as
\begin{eqnarray}
\chi(x_1\ldots
x_N)&=&{\mathrm
  e}^{-\frac{\mathrm{i}\kappa}{2}\left(w(x_i,x_j)+\sum_{x_l<x_m}^Nw(x_l,x_m)\right)}\left[\ldots
  A( \ldots k_i,k_j\ldots )\mathrm{e}^{\mathrm{i}(\ldots +k_ix_i+k_jx_j+\ldots )}\right.\nonumber\\
& &
\left.+A( \ldots k_j,k_i\ldots )\mathrm{e}^{\mathrm{i}(\ldots
  +k_jx_i+k_ix_j+\ldots )}+\ldots \right]\label{D1}
\end{eqnarray}
for the domain $0\le x_1<x_2< \ldots x_i<x_j \ldots <x_N \le L $ and
\begin{eqnarray}
\chi(x_1\ldots
x_N)&=&\mathrm
  {e}^{-\frac{\mathrm{i}\kappa}{2}\left(w(x_j,x_i)+\sum_{x_l<x_m}^Nw(x_l,x_m)\right)}
  \left[\ldots A( \ldots k_i,k_j\ldots )\mathrm{e}^{\mathrm{i}(\ldots +k_ix_j+k_jx_i+\ldots )}\right.\nonumber\\
& &
\left. \ldots +A( \ldots k_j,k_i\ldots )\mathrm{e}^{\mathrm{i}(\ldots
  +k_jx_j+k_ix_i+\ldots )}+\ldots \right]\label{D2}
\end{eqnarray}
for the domain $0\le x_1<x_2< \ldots x_j<x_i \ldots <x_N \le L $.
For $x_i=x_j$, the wavefunction is given by
\begin{eqnarray}
\chi(x_1\ldots x_N)&=&\mathrm{
  e}^{-\frac{\mathrm{i}\kappa}{2}\left(w(x_i,x_i)+\sum_{x_l<x_m}^Nw(x_l,x_m)\right)}
  \left[ \ldots A( \ldots k_i,k_j\ldots )\mathrm{e}^{\mathrm{i}(\ldots +(k_i+k_j)x_i+\ldots )}
\right.\nonumber\\
& &
\left.+A( \ldots k_j,k_i\ldots )\mathrm{e}^{\mathrm{i}(\ldots
  +(k_i+k_j)x_i+\ldots )}+\ldots \right].\label{D3}
\end{eqnarray}

Application of the derivative discontinuity condition (\ref{jump}) on
the wavefunctions (\ref{D1}),  (\ref{D2}) and  (\ref{D3}) gives
\begin{eqnarray}
& &\left\{ 
\mathrm{e}^{-\frac12 \mathrm{i} \kappa w(x_i,x_j)}+
\mathrm{e}^{-\frac12 \mathrm{i} \kappa w(x_j,x_i)}\right\} \mathrm{i}(k_j-k_i)\left[A(\ldots k_i,k_j\ldots)-
A(\ldots k_j,k_i\ldots)\right]\nonumber\\
&&=2c\left[A(\ldots k_i,k_j\ldots)+A(\ldots k_j,k_i\ldots) \right].
\end{eqnarray}
This equation gives a relation between the coefficients $A(k_{P1}\ldots k_{PN})$ 
of the form \cite{Kundu,BGO} 
\begin{equation}
A(\ldots k_j,k_i
\ldots)=\frac{k_j-k_i+\mathrm{i}c'}{k_j-k_i-\mathrm{i}c'}A(\ldots
k_i,k_j \ldots)
\end{equation}
which is the two-body scattering relation.  
Here the anyonic parameter $\kappa$ and the dynamical interaction $c$ are inextricably related
via the effective coupling constant \cite{Kundu,BGO}
\begin{equation}
c'=c/\cos(\kappa/2).
\end{equation}
This results in an interesting resonance-like effect in the effective
coupling constant $c'$ with respect to the statistical interaction around $\kappa = \pi$ \cite{BGO}.  
The single-valued eigenstate (\ref{state}) makes the continuity of the wavefunction (\ref{wave}) well
defined because the phase factors in the multi-valued wavefunction
(\ref{wave}) can be cancelled by those from permutations of the
particles in the eigenstate $\mid \!\! \Phi \rangle$ for different domains. 
The probability density can be written as $|\chi(x_1,\ldots,x_N)|^2$ which is subject to the scattering problem.
It suffices to request a continuity relation of the wavefunction of the form
\begin{equation}
\lim_{x_i=x_j}\chi(x_1,\ldots, x_i,x_j,\ldots, x_N)=\lim_{x_i=x_j}\chi(x_1,\ldots, x_j,x_i,\ldots, x_N)=\chi(x_1,\ldots, x_i,x_i,\ldots, x_N),
\end{equation}
or equivalently
\begin{eqnarray}
& &\lim_{x_i=x_j}\left[ A( \ldots k_i,k_j\ldots )\mathrm{e}^{\mathrm{i}(\ldots +k_ix_i+k_jx_j+\ldots )}
+A( \ldots k_j,k_i\ldots )\mathrm{e}^{\mathrm{i}(\ldots
  +k_jx_i+k_ix_j+\ldots )}+\ldots \right]\nonumber\\
   & & =\lim_{x_i=x_j}\left[ A( \ldots k_i,k_j\ldots )\mathrm{e}^{\mathrm{i}(\ldots +k_ix_j+k_jx_i+\ldots )}
+A( \ldots k_j,k_i\ldots )\mathrm{e}^{\mathrm{i}(\ldots+k_jx_j+k_ix_i+\ldots )}+\ldots \right]
\end{eqnarray}
which leads to an identity equation for the coefficients in the above two domains.

The  periodic boundary conditions $\chi(x_1=0,x_2\ldots x_N)= \chi(x_2 \ldots x_N,x_1=L)$ for the system  
are equivalent to $\chi(x_1,x_2\ldots x_N)=\mathrm{e}^{-\mathrm{i}\kappa(N-1)}\chi(x_1+L,x_2 \ldots x_N)$. 
The anyonic phases for $x_1$ at the left end and at the right end of the domain are different due
to the changes in the order of particles. 
Explicitly,
\begin{eqnarray}
 &&{\mathrm e}^{-\frac{\mathrm{i}\kappa}{2}\left(\sum_{i=2}^Nw(x_1,x_i)+\sum_{1<i<j}w(x_i,x_j)\right)}\sum_PA(k_{P1}\cdots k_{PN}){\mathrm e}^{\mathrm{i}(k_{P2}x_2\cdots
  +k_{PN}x_N)}\nonumber\\
&&= {\mathrm
  e}^{-\frac{\mathrm{i}\kappa}{2}\left(\sum_{i=2}^Nw(x_i,x_1)+
  \sum_{1<i<j}w(x_i,x_j) \right)}\sum_QA(k_{Q1}\cdots k_{QN}) {\mathrm e}^{\mathrm{i}(k_{Q1}x_2, k_{Q2}x_3\cdots
  +k_{Q(N-1)}x_N+k_{QN}L)}.\label{PBC}
\end{eqnarray}
Notice that the second sum in the phase factors in both sides will
cancel in deriving the Bethe ansatz equations below.
In order to compare the coefficients on both sides of the above equation,
we permute the order $Q$ back to the order $P$.
There are $N!$  patterns related to all different ways of displaying the positions of $k$'s.  
For example, in the right hand side of equation (\ref{PBC}) we can permute $k_1$ back from 
the right side of
$$
\left(\begin{array}{lllll} k_1&k_2&\ldots &k_{N-1}&k_N \\ k_2&k_3&\ldots &k_N &k_1 \end{array}\right)
$$
to the left side 
$$
\left(\begin{array}{lllll} k_1&k_2&\ldots &k_{N-1}&k_N \\  k_1&k_2&\ldots &k_{N-1} &k_N \end{array}\right).
$$
The Bethe Ansatz equations (BAE)
\begin{equation}
{\mathrm e}^{\mathrm{i}k_jL}=-{\mathrm e}^{\mathrm{i}\kappa(N-1)} \prod^N_{\ell = 1}
\frac{k_j-k_\ell+\mathrm{i}\, c'}{k_j-k_\ell-\mathrm{i}\, c'} 
\label{BA}
\end{equation}
follow directly through the scattering process. 
We demonstrate the derivation of the BAE (\ref{BA}) in Appendix \ref{APP3} for $N=3$.

It should be noted that this model reduces to the interacting Bose gas \cite{LL} for $\kappa=0$.
For $\kappa=\pi$ and $3\pi$ it reduces to the free Fermi gas.
The anyonic gas lies in the range $0\le \kappa \le \pi$ and $3\pi \le \kappa \le 4\pi$, where
the effective interaction $c'>0$.
However, if the anyon parameter $\kappa$ is tuned smoothly from
$\kappa <\pi$ to $\kappa>\pi$, i.e., $\pi \le\kappa \le 3\pi$, the
effective interaction is attractive. 
The ground state properties of this model have been studied in Ref.~\cite{BGO}.

\section{Thermodynamic Bethe Ansatz }
\label{sec:TBA}

In general the exponential phase factor in the BAE (\ref{BA}),
picked up from the statistical interaction during the scattering process,
may shift the system into higher excited states.
The total momentum is $p=N(N-1)\kappa/L+2d\pi/L$, where $d$ is an arbitrary integer.
This can be easily seen from the logarithmic form 
\begin{equation}
k_jL=2\pi I_j+\kappa(N-1)-2\sum_{l=1}^N\arctan (\frac{k_j-k_l}{c'}) \label{BAd}
\end{equation}
of the BAE, where $I_j=j-(N+1)/2$ with $j=1,\ldots, N$.
The energy is given by $E=\sum_{j=1}^Nk_j^2$.

In minimizing the energy we consider $\kappa(N-1) = \nu$ (mod $2\pi$) in the phase
factor with $-\pi \leq \nu \leq \pi$.
Each  quasimomentum $k_j$ shifts to $k_j+\nu/L$ in the ground state.
In general the anyonic factor $\kappa$ produces an effect like the
self-sustained Aharonov-Bohm flux resulting in metastable states.
In the thermodynamic limit, the BAE (\ref{BA}) can be written in the form of an integral equation
 \begin{equation}
\rho(k)=\frac{1}{2\pi}+\frac{1}{2\pi} \int_{-Q}^{Q} \frac{2c'}{c'^2+(k-k')^2}\rho(k')dk'.\label{BA-I}
\end{equation}
Here $Q$ is the cut-off momentum for the ground state.  
We have introduced the density of roots $\rho(k_j)=1/(L(k_{j+1}-k_j))$ in the interval $k_{j+1}-k_j$.  
It follows that in the thermodynamic limit the
lowest energy is given by $E=N(n^2e(\gamma,\kappa)+\nu^2/L^2)$ where
\begin{equation}
e(\gamma,\kappa)=\frac{\gamma^3}{\lambda^3}\int_{-1}^1g(x)x^2dx.
\end{equation}
The last term in the energy $E$ can be ignored compared to the kinetic energy.
The  root density $g(x) := \rho(Qx)$ and the parameter $\lambda =c/Q$
are determined by Lieb-Liniger type integral equations of the form
\begin{eqnarray}
 g(x)&=&\frac{1}{2\pi}+\frac{\lambda \cos({\kappa}/{2})
  }{\pi}\int_{-1}^{1}\frac{g(y)dy}{\lambda^2+\cos^2(\frac{\kappa}{2})(x-y)^2} \nonumber\\
  \lambda&=& \gamma \int_{-1}^1g(x)dx.
\label{BA2}
\end{eqnarray}
The numerical solution of this equation has been given in Ref.~\cite{BGO}.

At zero temperature, the quantum numbers $I_j$ form a uniform lattice from $-(N-1)/2$ to $(N-1)/2$. 
The quasimomenta $k_j$ form a non-uniform distribution $\rho(k)$ between $k=-Q$ and  $k=Q$, which is
determined by the BAE (\ref{BA-I}). 
The occupied lattice sites are called quasiparticles.  
For the excited states the lattice sites are not all occupied. 
This means that some $k$'s move out of the Dirac sea leaving some lattice sites unoccupied, 
the so called holes. 
If one $k$ leaves the Dirac sea, the remaining roots all move.
This collective behaviour phenomenon is governed by the linear
dispersion relation $E-E_0\approx v_c p$ as $p\to 0$, where the sound
velocity is $v_c\approx 2 n\pi(1-{4 \gamma^{-1} \cos(\kappa/2)})$ \cite{BGO}.
In this way, the density of quasiparticles and the density of the
holes are determined by the equation
\begin{equation}
\rho(k)+\rho_{\rm h}(k)=\frac{1}{2\pi}+\frac{1}{2\pi}\int
_{-\infty}^{\infty}\frac{2c'\rho(k')dk'}{c'^2+(k-k')^2}. \label{d-BA}
\end{equation}
The energy per particle for the state is given by
\begin{equation}
\frac{E}{N}=n^{-1}\int_{-\infty}^{\infty}\rho(k)k^2dk
\end{equation}
where the linear density $n=\int_{-\infty}^{\infty}\rho(k)dk$.

To obtain the thermodynamics following the Yang-Yang approach \cite{Y-Y}, 
an equilibrium state associated with the entropy
\begin{equation}
\frac{S}{N}=n^{-1}\int_{-\infty}^{\infty}\left[(\rho(k)+\rho_{\rm
    h}(k))\ln(\rho(k)+\rho_{\rm h}(k))-\rho(k)\ln\rho(k)-\rho_{\rm h}(k)\ln\rho_{\rm h}(k)\right]dk
\end{equation}
is a mixture of some eigenstates of the Hamiltonian.
For the equilibrium state the density of quasiparticles is determined by
minimizing the free energy $F=E-TS -\mu N$, i.e. $\delta F=0$, where $\mu$ is the
chemical potential.  
It follows that the TBA equation and the free energy are given by
\begin{equation}
\epsilon (k)=
\epsilon^0(k) -\mu-\frac{T}{2\pi}\int_{\infty}^{\infty}dk^{'}\theta
^{'}(k-k^{'})\ln(1+{\mathrm e}^{-\frac{\epsilon(k^{'})}{T}}) \label{TBA}
\end{equation}
\begin{equation}
\frac{F}{L} =\mu n-\frac{T}{2\pi}\int_{-\infty}^{\infty} dk
\ln(1+{\mathrm e}^{-\frac{\epsilon(k)}{T}}) \label{TBA-P}
\end{equation}
respectively. Here
\begin{equation}
\theta^{'}(x)=\frac{2c\cos(\kappa/2)}{c^2+\cos^2(\kappa/2)x^2}
\end{equation}
Here $\epsilon(k)$ is the dressed energy $e^{\epsilon(k)/T}:= \rho^h(k)/\rho(k)$ and the function
$\epsilon^0(k)=(k+\nu/L)^2$.
The thermodynamics of the system has been discussed in Ref.~\cite{BG}.
For $\kappa=0$, the thermodynamics of the 1D Bose gas has been studied
extensively, see, e.g. Refs~\cite{BG2,Bortz}.

The TBA equation (\ref{TBA}) also provides ground state properties and a clear picture  of the band filling. 
At zero temperature, the negative part $\epsilon^-$ of the dressed energy  makes a contribution to the free energy,
i.e. for $T\to 0$, the TBA equation  (\ref{TBA}) becomes
\begin{equation}
\epsilon(k)=k^2-\mu+\frac{1}{2\pi}\int_{-Q}^Q\frac{2c'\epsilon^-(k')}{c'^2+(k-k')^2}dk'.\label{TBA-B}
\end{equation}
The  pressure per unit length is given by $P_0=-\frac{1}{2\pi}\int_{-Q}^{Q}\epsilon^-(k)dk$.  
For the Tonks-Girardeau regime, i.e.  for $\gamma \gg 1$, we have
\begin{equation}
\epsilon(k)\approx k^2-\mu-\frac{2P_0c'}{c'^2+k^2}.
\end{equation}
{}From this equation, with the help of the conditions $\epsilon(Q)=0$ and
$n=\partial P_0/\partial \mu$,
one can obtain physical quantities, such as the ground state energy $E_0$, chemical
potential $\mu$, pressure $P_0$  and  the cut-off momentum $Q$.
The results are
\begin{eqnarray}
&&E_0/L\approx \frac{1}{3}n^3\pi^2\left(1-\frac{4\cos(\kappa/2)}{\gamma}\right),\,\,\,\,
\mu \approx n^2\pi^2 \left(1-\frac{16\cos(\kappa/2)}{3\gamma}\right),\nonumber\\
&&P_0\approx \frac{2}{3}n^3\pi^2\left(1-\frac{6\cos(\kappa/2)}{\gamma}\right),\,\,\,\,
Q \approx n\pi \left(1-\frac{2\cos(\kappa/2)}{\gamma}\right).\label{TBA-r}
\end{eqnarray}
The macroscopic velocity is
\begin{equation}
v=\sqrt{2\frac{\partial  P_0}{\partial n}}\approx 2 n\pi(1-{4 \gamma^{-1} \cos(\kappa/2)}).
\end{equation}
On the other hand, from the BAE (\ref{BA-I}), we have
\begin{equation}
\rho(k)\approx \frac{1}{2\pi}\left(1+\frac{2nc'}{c'^2+k^2}+\frac{4c'kp}{(c'^2+k^2)^2}\right), 
\end{equation}
where $p$ is the total momentum. 
Then from the condition
$n=\int_{-Q}^{Q}\rho(k)dk$, we have
\begin{eqnarray}
& &Q \approx n\pi \left(1-\frac{2\cos(\kappa/2)}{\gamma}\right),\,\,\,\,
E_0/L\approx
\frac{1}{3}n^3\pi^2 \left(1-\frac{4\cos(\kappa/2)}{\gamma}+\frac{12\cos^2(\kappa/2)}{\gamma^2}\right)
\end{eqnarray}
which coincide with the TBA results (\ref{TBA-r}).

\section{conclusion}
\label{conclusion}

In this paper we have derived the Bethe Ansatz solution of the 1D $\delta$-function
interacting model of anyons, which is a natural extension of the 1D $\delta$-function 
interacting Bose gas, recovered in the limit $\kappa=0$.
The ground state properties (\ref{TBA-r}) have been derived from the TBA equations (\ref{TBA}). 
The subtlety of the anyonic symmetry results in rich quantum effects for the degenerate anyon
gas \cite{BGO,BG}. 
Here we remark that a generalization of this model to the 1D $\delta$-function interacting model of 
anyons with internal spin degrees of freedom should be possible. 
We expect that the combination of anyonic statistical interaction, dynamical interaction and spin degrees 
will result in new quantum effects.

\section*{Acknowledgements}
This work has been supported by the Australian Research Council.  The
authors thank M. Bortz, N. Oelkers and H.-T. Yang for helpful
discussions.  XWG thanks the Department of Mathematics, University of
Science and Technology of China for their kind hospitality.  JSH is
supported by the NSF of China under Grant No. 10301030.

\clearpage

\appendix
\section{Three particle eigenstate} \label{APPa}

Here we give the explicit eigenstate (\ref{state}) for  $N=3$.
For the order of particle coordinates $x_1<x_2<x_3$ the wavefunction is
\begin{eqnarray}
\chi(x_1,x_2,x_3)&=&\mathrm{e}^{-\mathrm{i}\frac{\kappa}{2}(w(x_1,x_2)+w(x_1,x_3)+w(x_2,x_3))}
\sum_PA(k_{P1},k_{P2},k_{P3})\mathrm{e}^{k_{P1}x_1+k_{P2}x_2+k_{P3}x_3}. \label{M1}
\end{eqnarray}
The sum runs over all $3!$ permutations $P$.
According to the definition in the anyonic commutation
relations the above phase factor is ${\rm e}^{\mathrm{i}\frac{3}{2}\kappa}$ for $N=3$ in the 
assigned order $x_1<x_2<x_3$.
In this domain the eigenstate for hamiltonian (\ref{Ham1}) is
\begin{eqnarray}
|\Phi\rangle& =&\int dx^3
 \mathrm{e}^{-\mathrm{i}\frac{3\kappa}{2}}\mathrm{e}^{-\mathrm{i}\frac{\kappa}{2}(w(x_1,x_2)+w(x_1,x_3)+w(x_2,x_3))}
 \left[\sum_P A(k_{p1},k_{p2},k_{p3})
 \mathrm{e}^{k_{P1}x_1+k_{P2}x_2+k_{P3}x_3}\right]\nonumber\\
&&\times \Psi^{\dagger}(x_1)
 \Psi^{\dagger}(x_2)\Psi^{\dagger}(x_3)|0\rangle\nonumber\\
&=&\int dx^3\left[\sum_PA(k_{p1},k_{p2},k_{p3})e^{k_{P1}x_1+k_{P2}x_2+k_{P3}x_3}\right] \Psi^{\dagger}(x_1) \Psi^{\dagger}(x_2)\Psi^{\dagger}(x_3)|0\rangle.\label{w1}
\end{eqnarray}
The phase factor is unity in the assigned order $x_1<x_2<x_3$.

Now consider two-particle exchange, i.e. the particle order $x_2<x_1<x_3$.  
We can write the wavefunction as
\begin{eqnarray}
\chi(x_2,x_1,x_3)&=&\mathrm{e}^{-\mathrm{i}\frac{\kappa}{2}(w(x_2,x_1)+w(x_2,x_3)+w(x_1,x_3))}\left[\sum_P
A(k_{P1},k_{P2},k_{P3})\mathrm{e}^{k_{P1}x_2+k_{P2}x_1+k_{P3}x_3}\right].\label{M2}
\end{eqnarray}
Thus the eigenstate is given by
\begin{eqnarray}
|\Phi\rangle& =&\int dx^3
 \mathrm{e}^{-\mathrm{i}\frac{3\kappa}{2}}\mathrm{e}^{-\mathrm{i}\frac{\kappa}{2}(w(x_2,x_1)+w(x_2,x_3)+w(x_1,x_3))}\left[\sum_P
 A(k_{P1},k_{P2},k_{P3})\mathrm{e}^{k_{P1}x_2+k_{P2}x_1+k_{P3}x_3}\right]\nonumber
 \\
 &&\times \Psi^{\dagger}(x_2)
 \Psi^{\dagger}(x_1)\Psi^{\dagger}(x_3)|0\rangle\nonumber \\
&= &\int dx^3
 \mathrm{e}^{-\mathrm{i}\frac{3\kappa}{2}}\mathrm{e}^{-\mathrm{i}\frac{\kappa}{2}(w(x_1,x_2)+w(x_2,x_3)+w(x_1,x_3))}\left[\sum_P
 A(k_{P1},k_{P2},k_{P3})\mathrm{e}^{k_{P1}x_2+k_{P2}x_1+k_{P3}x_3}\right]\nonumber
 \\
&&\times \Psi^{\dagger}(x_1)
 \Psi^{\dagger}(x_2)\Psi^{\dagger}(x_3)|0\rangle \nonumber \\
&=& \int dx^3\left[\sum_PA(k_{P1},k_{P2},k_{P3}))e^{k_{P1}x_2+k_{P2}x_1+k_{P3}x_3}\right]\Psi^{\dagger}(x_1) \Psi^{\dagger}(x_2)\Psi^{\dagger}(x_3)|0\rangle.
 \label{w4}
\end{eqnarray}
It is clear that the summation in the phase factor of the wavefunction
 $\chi$ simply counts the changes of the order in the relevent domain.
 

\section{From quantum field to many-body quantum mechanics: $N=3$}\label{APP1}

In this appendix we explicitly demonstrate the relation between the quantum field theoretic 
hamiltonian and the quantum mechanics many-body problem.
First, we apply the periodic conditions and integrate by
parts in the first part of hamiltonian (\ref{Ham1}) 
\begin{equation}\label{HP1}
\int\limits_0^{L}dx [\partial_x\Psi^\dag(x)][
\partial_x\Psi(x)]=-\int\limits_0^{L}dx(\partial^2_x\Psi^\dag(x))\Psi(x).
\end{equation}
For $N=3$, the $N$-particle eigenstate $\mid\!\Phi\rangle$ in (\ref{state}) reads
\begin{equation}
\mid\!\Phi\rangle =\int\limits_0^Ldx_1dx_2dx_3{\mathrm e}^{-{\mathrm
i}\frac{3}{2}\kappa}\chi(x_1,x_2,x_3)\Psi^\dag(x_1)\Psi^\dag(x_2)\Psi^\dag(x_3)
\! \mid \! 0\rangle, \quad x_1<x_2<x_3. \label{threeparticlestate}
\end{equation}
We consider operations $H$ on this state. 
Let $A=\int\limits_0^{L}dx
[\partial_x\Psi^\dag(x)][\partial_x\Psi(x)] \mid\!\Phi\rangle$, then
\begin{eqnarray}
A\!\!&\!=\!&\!-\!\!\int\limits_0^L\!dxdx_1dx_2dx_3{\mathrm
e}^{-{\mathrm
i}\frac{3}{2}\kappa}\chi(x_1,x_2,x_3)(\partial^2_x\Psi^\dag(x))\Psi(x)
\Psi^\dag(x_1)\Psi^\dag(x_2)\Psi^\dag(x_3) \! \mid \! 0\rangle
\nonumber\\
\!\!&\!\mbox{\hspace{-1.cm}}=\!&\mbox{\hspace{-0.6cm}}-\!\!\!\int\limits_0^L\!\!dxdx_1dx_2dx_3{\mathrm
e}^{-{\mathrm
i}\frac{3}{2}\kappa}\chi(x_1,x_2,x_3)(\partial^2_x\Psi^\dag(x))
\left({\mathrm e}^{-{\mathrm i}\kappa w(x,x_1) }
\Psi^\dag(x_1)\Psi(x)+\delta(x-x_1) \right)
\Psi^\dag(x_2)\Psi^\dag(x_3) \! \mid \! 0\rangle
 \nonumber\\
\!\!&\!\mbox{\hspace{-1.cm}}=\!&\mbox{\hspace{-0.6cm}}-\!\!\!\int\limits_0^L\!\!dxdx_1dx_2dx_3{\mathrm
e}^{-{\mathrm
i}\frac{3}{2}\kappa}\chi(x_1,x_2,x_3)(\partial^2_x\Psi^\dag(x))
\delta(x-x_1) \Psi^\dag(x_2)\Psi^\dag(x_3) \! \mid \! 0\rangle
 \nonumber\\
\!\!&\!\mbox{\hspace{-1.cm}}\mbox{}\!&\mbox{\hspace{-0.6cm}}-\!\!\!\int\limits_0^L\!\!dxdx_1dx_2dx_3{\mathrm
e}^{-{\mathrm
i}\frac{3}{2}\kappa}\chi(x_1,x_2,x_3)(\partial^2_x\Psi^\dag(x))
{\mathrm e}^{-{\mathrm i}\kappa w(x,x_1) } \Psi^\dag(x_1)\Psi(x)
\Psi^\dag(x_2)\Psi^\dag(x_3) \! \mid \!
0\rangle.\label{firststepexchange}
\end{eqnarray}
Here we succeeded in exchanging the position for $\Psi(x)$ and
$\Psi^{\dag}(x_1)$ by means of the commutation relations (\ref{comm}). 
In (\ref{firststepexchange}), the first term is the right term.   
We continue this procedure for the second term until $\Psi(x)$ is moved to the right
hand side of $\Psi^{\dag}(x_3)$. 
Thus
\begin{eqnarray}
A\!\!&\!=\!&-\!\!\!\int\limits_0^L\!\!dxdx_1dx_2dx_3{\mathrm
e}^{-{\mathrm
i}\frac{3}{2}\kappa}\chi(x_1,x_2,x_3)(\partial^2_x\Psi^\dag(x))
\delta(x-x_1) \Psi^\dag(x_2)\Psi^\dag(x_3) \! \mid \! 0\rangle
 \nonumber\\
\!\!&\!\mbox{\hspace{-2.5cm}}\mbox{}\!&\mbox{\hspace{-1.2cm}}-\!\!\!\!
\int\limits_0^L\!\!\!dxdx_1dx_2dx_3{\mathrm e}\!^{-{\mathrm
i}(\frac{3}{2}\kappa +\kappa
w(x,x_1))}\chi\!(x_1\!,\!x_2\!,\!x_3\!)\!(\partial^2_x\Psi^\dag\!(x)\!)\!
\Psi^\dag\!(x_1\!)\!\!\left(\!{\mathrm e}\!^{-{\mathrm i}\kappa
w(x,x_2)}\!
\Psi^\dag\!(x_2\!)\!\Psi\!(x)\!+\!\delta(x\!\!-\!\!x_2)\!\right)\!\!\Psi^\dag\!(x_3)\!\!
\mid\!\!0\!\rangle\nonumber\\
&=&-\!\!\!\int\limits_0^L\!\!dxdx_1dx_2dx_3{\mathrm e}^{-{\mathrm
i}\frac{3}{2}\kappa}\chi(x_1,x_2,x_3)(\partial^2_x\Psi^\dag(x))
\delta(x-x_1) \Psi^\dag(x_2)\Psi^\dag(x_3) \! \mid \! 0\rangle
 \nonumber\\
\!\!&\!\mbox{}\!&-\!\!\!\!
\int\limits_0^L\!\!\!dxdx_1dx_2dx_3{\mathrm e}\!^{-{\mathrm
i}(\frac{3}{2}\kappa +\kappa
w(x,x_1))}\chi\!(x_1\!,\!x_2\!,\!x_3\!)\!(\partial^2_x\Psi^\dag\!(x)\!)\!
\Psi^\dag\!(x_1\!)\delta(x\!-\!x_2)\Psi^\dag\!(x_3)\!\!
\mid\!\!0\!\rangle\nonumber\\
\!\!&\!\mbox{\hspace{-2.5cm}}\mbox{}\!&\mbox{\hspace{-1.2cm}}-\!\!\!\!
\int\limits_0^L\!\!\!dxdx_1dx_2dx_3{\mathrm e}\!^{-{\mathrm
i}(\frac{3}{2}\kappa +\kappa w(x,x_1)+\kappa
w(x,x_2))}\chi\!(x_1\!,\!x_2\!,\!x_3\!)\!(\partial^2_x\Psi^\dag\!(x)\!)\!
\Psi^\dag\!(x_1\!)\!\!\left(
\Psi^\dag\!(x_2\!)\!\Psi\!(x)\!\!\right)\!\!\Psi^\dag\!(x_3)\!\!
\mid\!\!0\!\rangle\nonumber\\
&=&-\!\!\!\int\limits_0^L\!\!dxdx_1dx_2dx_3{\mathrm e}^{-{\mathrm
i}\frac{3}{2}\kappa}\chi(x_1,x_2,x_3)(\partial^2_x\Psi^\dag(x))
\delta(x-x_1) \Psi^\dag(x_2)\Psi^\dag(x_3) \! \mid \! 0\rangle
 \nonumber\\
\!\!&\!\mbox{}\!&-\!\!\!\!
\int\limits_0^L\!\!\!dxdx_1dx_2dx_3{\mathrm e}\!^{-{\mathrm
i}(\frac{3}{2}\kappa +\kappa
w(x,x_1))}\chi\!(x_1\!,\!x_2\!,\!x_3\!)\!(\partial^2_x\Psi^\dag\!(x)\!)\!
\Psi^\dag\!(x_1\!)\delta(x\!-\!x_2)\Psi^\dag\!(x_3)\!\!
\mid\!\!0\!\rangle\nonumber\\
\!\!&\!\mbox{\hspace{-2.5cm}}\mbox{}\!&\mbox{\hspace{-0.8cm}}-\!\!\!\!
\int\limits_0^L\!\!\!dxdx_1dx_2dx_3{\mathrm e}\!^{-{\mathrm
i}(\frac{3}{2}\kappa +\kappa w(x,x_1)+\kappa
w(x,x_2))}\chi\!(x_1\!,\!x_2\!,\!x_3\!)\!(\partial^2_x\Psi^\dag\!(x)\!)\!
\Psi^\dag\!(x_1\!) \Psi^\dag\!(x_2\!)\delta(x\!\!-\!\!x_3)\!\!
\mid\!\!0\!\rangle. \nonumber
\end{eqnarray}
The results $\Psi (x)\Psi ^{\dagger}(x_3)={\mathrm e}^{-\mathrm{i}\kappa
  w(x,x_3)}\Psi ^{\dagger}(x_3)\Psi(x)+\delta(x-x_3)$ and $\Psi(x)\mid 0\rangle=0$ 
  are used in the last step. 
Using integration by parts and the integral property of the $\delta$-function, we then have
\begin{eqnarray}
A&=&-\!\!\!\int\limits_0^L\!\!dx_1dx_2dx_3{\mathrm e}^{-{\mathrm
i}\frac{3}{2}\kappa} (\partial_{x_1}^2\chi(x_1,x_2,x_3))
\Psi^\dag(x_1)\Psi^\dag(x_2)\Psi^\dag(x_3) \! \mid \! 0\rangle
 \nonumber\\
\!\!&\!\mbox{}\!&-\!\!\!\! \int\limits_0^L\!\!\!dx_1dx_2dx_3{\mathrm
e}\!^{-{\mathrm i}(\frac{3}{2}\kappa +\kappa
w(x_2,x_1))}(\partial_{x_2}^2\chi\!(x_1\!,\!x_2\!,\!x_3\!))\!\Psi^\dag(x_2)\!
\Psi^\dag\!(x_1\!)\Psi^\dag\!(x_3)\!\!
\mid\!\!0\!\rangle\nonumber\\
\!\!&\!\mbox{\hspace{-2.5cm}}\mbox{}\!&\mbox{\hspace{-0.8cm}}-\!\!\!\!
\int\limits_0^L\!\!\!dx_1dx_2dx_3{\mathrm e}\!^{-{\mathrm
i}(\frac{3}{2}\kappa +\kappa w(x_3,x_1)+\kappa
w(x_3,x_2))}(\partial_{x_3}^2\chi\!(x_1\!,\!x_2\!,\!x_3\!))\Psi^\dag\!(x_3)
\Psi^\dag\!(x_1\!) \Psi^\dag\!(x_2\!) \mid\!\!0\!\rangle.\nonumber \\
&=&-\int\limits_0^L\!\!dx_1dx_2dx_3{\mathrm e}^{-{\mathrm
i}\frac{3}{2}\kappa} \left((\partial_{x_1}^2+\partial_{x_2}^2
+\partial_{x_3}^2 )\chi(x_1,x_2,x_3)\right)
\Psi^\dag(x_1)\Psi^\dag(x_2)\Psi^\dag(x_3) \! \mid \! 0\rangle.
\label{firstpart}
\end{eqnarray}
Here the commutation relations (\ref{comm}) are used to form the pattern
$\Psi^{\dag}(x_1)\Psi^{\dag}(x_2)\Psi^{\dag}(x_3)$ by moving around
$\Psi^{\dag}(x_2)$ and $\Psi^{\dag}(x_3)$.

We now turn to the second part of the hamiltonian (\ref{Ham1}).
Accordingly, let 
$B=\int\limits_0^Ldx \Psi^{\dag}(x)\Psi^{\dag}(x)\Psi(x)\Psi(x)\mid \Phi\rangle$, then
\begin{eqnarray}
B&=&\int\limits_0^Ldxdx_1dx_2dx_3 {\mathrm e}^{-{\mathrm
i}\frac{3}{2}\kappa}\chi(x_1,x_2,x_3)\Psi^{\dag}(x)\Psi^{\dag}(x)\Psi(x)\Psi(x)
\Psi^\dag(x_1)\Psi^\dag(x_2)\Psi^\dag(x_3) \! \mid \! 0\rangle \nonumber\\
&\mbox{\hspace{-2cm}}=&\mbox{-\hspace{-1.5cm}}\int\limits_0^L\!dxdx_1dx_2dx_3
{\mathrm e}^{-{\mathrm
i}\frac{3}{2}\kappa}\chi(x_1\!,\!x_2\!,\!x_3)\!(\Psi^{\dag}(x))^2\Psi(x)
\!\left( \Psi^{\dag}(x_1) \Psi(x){\mathrm e}^{-{\mathrm i}\kappa
w(x,x_1)}\!+\!\delta(x\!-\!x_1)
 \right)\!\Psi^\dag(x_2)\Psi^\dag(x_3) \!
\mid \! 0\rangle \nonumber\\
&=& \int\limits_0^Ldxdx_1dx_2dx_3 {\mathrm e}^{-{\mathrm
i}(\frac{3}{2}\kappa+\kappa
w(x,x_1))}\chi(x_1,x_2,x_3)(\Psi^{\dag}(x))^2\Psi(x)
\Psi^{\dag}(x_1) \Psi(x) \Psi^\dag(x_2)\Psi^\dag(x_3) \!
\mid \! 0\rangle \nonumber\\
&+& \int\limits_0^Ldxdx_1dx_2dx_3 {\mathrm e}^{-{\mathrm
i}\frac{3}{2}\kappa}\chi(x_1,x_2,x_3)(\Psi^{\dag}(x))^2\Psi(x)\delta(x-x_1)
\Psi^\dag(x_2)\Psi^\dag(x_3) \!
\mid \! 0\rangle \nonumber\\
&\mbox{\hspace{-3cm}}=&\mbox{\hspace{-1.7cm}}\!\!\int\limits_0^L\!\!dxdx_1dx_2dx_3
{\mathrm e}\!^{-{\mathrm i}(\frac{3}{2}\kappa+\kappa
w(x,x_1))}\chi(x_1,\!x_2,\!x_3)\!(\Psi^{\dag}(x))^2\Psi(x)
\Psi^{\dag}(x_1)\!\left(\!
 \Psi^\dag\!(x_2)\Psi(x){\mathrm e}^{-{\mathrm i}\kappa w(x,x_2)}
\!\!+\!\delta(x\!-\!x_2)\!\right)\!\!\Psi^\dag\!(x_3) \!
\!\mid\!\!0\!\rangle \nonumber\\
&+& \int\limits_0^Ldxdx_1dx_2dx_3 {\mathrm e}^{-{\mathrm
i}\frac{3}{2}\kappa}\chi(x_1,x_2,x_3)(\Psi^{\dag}(x_1))^2\Psi(x_1)
\Psi^\dag(x_2)\Psi^\dag(x_3) \!
\mid \! 0\rangle \nonumber\\
&\mbox{\hspace{-2cm}}=&\mbox{\hspace{-1.2cm}}\!\!\int\limits_0^L\!\!dxdx_1dx_2dx_3
{\mathrm e}\!^{-{\mathrm i}(\frac{3}{2}\kappa+\kappa
w(x,x_1))+\kappa
w(x,x_2)}\chi(x_1,\!x_2,\!x_3)\!(\Psi^{\dag}(x))^2\Psi(x)
\Psi^{\dag}(x_1)\!
 \Psi^\dag\!(x_2)\Psi(x)\Psi^\dag\!(x_3) \!
\mid \! 0\rangle \nonumber\\
&\mbox{\hspace{-2cm}}+&\mbox{\hspace{-1.2cm}}\!\!\int\limits_0^L\!\!dxdx_1dx_2dx_3
{\mathrm e}\!^{-{\mathrm i}(\frac{3}{2}\kappa+\kappa
w(x,x_1))}\chi(x_1,\!x_2,\!x_3)\!(\Psi^{\dag}(x))^2\Psi(x)
\Psi^{\dag}(x_1)\delta(x\!-\!x_2)\Psi^\dag\!(x_3) \!
\mid \! 0\rangle \nonumber\\
&\mbox{\hspace{-2cm}}+&\mbox{\hspace{-1.2cm}}\int\limits_0^Ldxdx_1dx_2dx_3
{\mathrm e}^{-{\mathrm
i}\frac{3}{2}\kappa}\chi(x_1,x_2,x_3)(\Psi^{\dag}(x_1))^2 \left(
\Psi^\dag(x_2)\Psi(x_1){\mathrm e}^{-{\mathrm i}\kappa w(x_1,x_2)}+
\delta(x_1-x_2)
 \right)
 \Psi^\dag(x_3) \!
\mid \! 0\rangle \nonumber\\
&\mbox{\hspace{-2cm}}=&\mbox{\hspace{-1.2cm}}\!\!\int\limits_0^L\!\!dxdx_1dx_2dx_3
{\mathrm e}\!^{-{\mathrm i}(\frac{3}{2}\kappa+\kappa
w(x,x_1))+\kappa
w(x,x_2)}\chi(x_1,\!x_2,\!x_3)\!(\Psi^{\dag}(x))^2\Psi(x)
\Psi^{\dag}(x_1)\!
 \Psi^\dag\!(x_2)
\delta(x-x_3)
  \!\mid \! 0\rangle \nonumber\\
&\mbox{\hspace{-2cm}}+&\mbox{\hspace{-1.2cm}}\!\!\int\limits_0^L\!\!dx_1dx_2dx_3
{\mathrm e}\!^{-{\mathrm i}(\frac{3}{2}\kappa+\kappa
w(x_2,x_1))}\chi(x_1,\!x_2,\!x_3)\!(\Psi^{\dag}(x_2))^2\Psi(x_2)
\Psi^{\dag}(x_1)\Psi^\dag\!(x_3) \!
\mid \! 0\rangle \nonumber\\
&\mbox{\hspace{-2cm}}+&\mbox{\hspace{-1.2cm}}\int\limits_0^Ldxdx_1dx_2dx_3
{\mathrm e}^{-{\mathrm i}(\frac{3}{2}\kappa+\kappa
w(x_1,x_2))}\chi(x_1,x_2,x_3)(\Psi^{\dag}(x_1))^2
\Psi^\dag(x_2)\Psi(x_1)
 \Psi^\dag(x_3) \!
\mid \! 0\rangle \nonumber\\
&\mbox{\hspace{-2cm}}+&\mbox{\hspace{-1.2cm}}\int\limits_0^Ldxdx_1dx_2dx_3
{\mathrm e}^{-{\mathrm
i}\frac{3}{2}\kappa}\chi(x_1,x_2,x_3)(\Psi^{\dag}(x_1))^2
\delta(x_1-x_2)
 \Psi^\dag(x_3) \!
\mid \! 0\rangle \nonumber\\
&\mbox{\hspace{-2cm}}=&\mbox{\hspace{-1.2cm}}\!\!\int\limits_0^L\!\!dx_1dx_2dx_3
{\mathrm e}\!^{-{\mathrm i}(\frac{3}{2}\kappa+\kappa
w(x_3,x_1))+\kappa
w(x_3,x_2)}\chi(x_1,\!x_2,\!x_3)\!(\Psi^{\dag}(x_3))^2\Psi(x_3)
\Psi^{\dag}(x_1)\!
 \Psi^\dag\!(x_2)
  \!\mid \! 0\rangle \nonumber\\
&\mbox{\hspace{-2cm}}+&\mbox{\hspace{-1.2cm}}\!\!\int\limits_0^L\!\!dx_1dx_2dx_3
{\mathrm e}\!^{-{\mathrm i}(\frac{3}{2}\kappa+\kappa
w(x_2,x_1))}\chi(x_1,\!x_2,\!x_3)\!(\Psi^{\dag}(x_2))^2\Psi(x_2)
\Psi^{\dag}(x_1)\Psi^\dag\!(x_3) \!
\mid \! 0\rangle \nonumber\\
&\mbox{\hspace{-2cm}}+&\mbox{\hspace{-1.4cm}}\int\limits_0^L\!\!dx_1dx_2dx_3
{\mathrm e}\!^{-{\mathrm i}(\frac{3}{2}\kappa+\kappa
w(x_1,x_2))}\chi(x_1\!,x_2\!,x_3)\!(\Psi^{\dag}\!(x_1))^2
\Psi^\dag\!(x_2)\!\!\left(\Psi^{\dag}\!(x_3)\Psi(x_1){\mathrm
e}\!^{-{\mathrm i}\kappa w(x_1,x_3)}\!
+\!\delta(x_1\!-\!x_3)\!\right)\!\!\mid\!0\!\rangle \nonumber\\
&\mbox{\hspace{-2cm}}+&\mbox{\hspace{-1.2cm}}\int\limits_0^Ldx_1dx_2dx_3
{\mathrm e}^{-{\mathrm
i}\frac{3}{2}\kappa}\chi(x_1,x_2,x_3)(\Psi^{\dag}(x_1))^2
\delta(x_1-x_2)
 \Psi^\dag(x_3) \!\!
\mid \! 0\!\rangle .\label{secondpartonestep}
\end{eqnarray}
Denote these four last terms by $b_4$, $b_3$, $b_2$ and $b_1$ 
in order from top to bottom.

We now discuss these terms case by case. 
The simplest case is $b_1$.
With the help of the property of the $\delta$ function it becomes
\begin{equation}\label{secondpartb1}
b_1=\int\limits_0^Ldx_1dx_2dx_3 {\mathrm e}^{-{\mathrm
i}\frac{3}{2}\kappa}\chi(x_1,x_2,x_3)\delta(x_1-x_2)\Psi^{\dag}(x_1)\Psi^{\dag}(x_2)\Psi^{\dag}(x_3)\mid
\!0\rangle
\end{equation}
as expected.  
By using $\Psi(x)\mid\! 0\rangle=0$, we have
\begin{eqnarray}
b_2 &=&\int\limits_0^L\!\!dx_1dx_2dx_3 {\mathrm e}\!^{-{\mathrm
i}(\frac{3}{2}\kappa+\kappa
w(x_1,x_2))}\chi(x_1\!,x_2\!,x_3)\delta(x_1\!-\!x_3)(\Psi^{\dag}\!(x_1))^2
\Psi^\dag\!(x_2)\!\mid\!0\!\rangle\nonumber\\
 &=&\int\limits_0^L\!\!dx_1dx_2dx_3 {\mathrm e}\!^{-{\mathrm
i}(\frac{3}{2}\kappa+\kappa
w(x_1,x_2))}\chi(x_1\!,x_2\!,x_3)\delta(x_1\!-\!x_3)\Psi^{\dag}\!(x_1)
\Psi^{\dag}\!(x_3)\Psi^\dag\!(x_2)\!\mid\!0\!\rangle\nonumber\\
&=&\int\limits_0^L\!\!dx_1dx_2dx_3 {\mathrm e}\!^{-{\mathrm
i}(\frac{3}{2}\kappa+\kappa
w(x_1,x_2))}\chi(x_1\!,x_2\!,x_3)\delta(x_1\!-\!x_3)\Psi^{\dag}\!(x_1)
{\mathrm e}^{{\mathrm i}\kappa w(x_3,x_2)}\Psi^\dag\!(x_2)\Psi^{\dag}\!(x_3)\!\mid\!0\!\rangle\nonumber\\
&=&\int\limits_0^L\!\!dx_1dx_2dx_3 {\mathrm e}\!^{-{\mathrm
i}\frac{3}{2}\kappa}\chi(x_1,x_2,x_3)\delta(x_1\!-\!x_3)\Psi^{\dag}\!(x_1)
\Psi^\dag\!(x_2)\Psi^{\dag}\!(x_3)\!\mid\!0\!\rangle.\label{secondpartb2}
\end{eqnarray}
The term $b_3$ is more complicated because we have to move
$\Psi(x_2)$ to the right hand side of $\Psi^{\dag}(x_3)$, namely
\begin{eqnarray}
\mbox{\hspace{-2cm}}b_3&\!\!=&\!\!\!\!\int\limits_0^L\!\!dx_1dx_2dx_3
{\mathrm e}\!^{-{\mathrm i}\kappa (\frac{3}{2}+
w(x_2,x_1))}\chi(x_1,\!x_2,\!x_3)\!(\Psi^{\dag}(x_2))^2\!\left(\Psi^{\dag}(x_1)\Psi(x_2)
{\mathrm e}^{-{\mathrm i}\kappa w(x_2,x_1) }\!+\delta(x_1\!-\!x_2)
\right)\!\Psi^\dag\!(x_3) \! \mid \! 0\rangle \nonumber\\
\mbox{\hspace{-2cm}}&\!\!=&\!\!\!\!\int\limits_0^L\!\!dx_1dx_2dx_3
{\mathrm e}\!^{-{\mathrm i}(\frac{3}{2}\kappa+2\kappa
w(x_2,x_1))}\chi(x_1,\!x_2,\!x_3)\!(\Psi^{\dag}(x_2))^2
\Psi^{\dag}(x_1)\Psi(x_2) \Psi^\dag\!(x_3) \! \mid \! 0\rangle
\nonumber\\
\mbox{\hspace{-2cm}}&\!\!+&\!\!\!\!\int\limits_0^L\!\!dx_1dx_2dx_3
{\mathrm e}\!^{-{\mathrm i}(\frac{3}{2}\kappa+\kappa
w(x_2,x_1))}\chi(x_1,\!x_2,\!x_3)\!(\Psi^{\dag}(x_2))^2\delta(x_1\!-\!x_2)\Psi^\dag\!(x_3)
\! \mid \! 0\rangle \nonumber\\
\mbox{\hspace{-2cm}}&\!\!=&\!\!\!\!\int\limits_0^L\!\!dx_1dx_2dx_3
{\mathrm e}\!^{-{\mathrm i}(\frac{3}{2}\kappa+2\kappa
w(x_2,x_1))}\chi(x_1,\!x_2,\!x_3)\!(\Psi^{\dag}(x_2))^2
\Psi^{\dag}(x_1)\! \left(\Psi^\dag\!(x_3)\Psi(x_2){\mathrm
e}^{-{\mathrm i}\kappa w(x_2,x_3)}\!+\delta(x_2\!-\!x_3)\right)\!
\!\mid \! 0\rangle
\nonumber\\
\mbox{\hspace{-2cm}}&\!\!+&\!\!\!\!\int\limits_0^L\!\!dx_1dx_2dx_3
{\mathrm e}\!^{-{\mathrm
i}\frac{3}{2}\kappa}\chi(x_1,\!x_2,\!x_3)\delta(x_1\!-\!x_2)\Psi^{\dag}(x_1)\Psi^{\dag}(x_2)\Psi^\dag\!(x_3)
\! \mid \! 0\rangle \nonumber\\
\mbox{\hspace{-2cm}}&\!\!=&\!\!\!\!\int\limits_0^L\!\!dx_1dx_2dx_3
{\mathrm e}\!^{-{\mathrm i}(\frac{3}{2}\kappa+2\kappa
w(x_2,x_1))}\chi(x_1,\!x_2,\!x_3)\!\Psi^{\dag}(x_2)\Psi^{\dag}(x_3)
\Psi^{\dag}(x_1)\delta(x_2-x_3)\mid \! 0\rangle
\nonumber\\
\mbox{\hspace{-2cm}}&\!\!+&\!\!\!\!\int\limits_0^L\!\!dx_1dx_2dx_3
{\mathrm e}\!^{-{\mathrm
i}\frac{3}{2}\kappa}\chi(x_1,\!x_2,\!x_3)\delta(x_1\!-\!x_2)\Psi^{\dag}(x_1)\Psi^{\dag}(x_2)\Psi^\dag\!(x_3)
\! \mid \! 0\rangle.\nonumber
\end{eqnarray}
Now using commutation relations (\ref{comm}), we move
$\Psi^{\dag}(x_1)$ to the left side of $\Psi^{\dag}(x_2)$ such that
\begin{eqnarray}
\mbox{\hspace{-2cm}}b_3&\!\!=&\!\!\!\!\int\limits_0^L\!\!dx_1dx_2dx_3
{\mathrm e}\!^{-{\mathrm
i}\frac{3}{2}\kappa}\chi(x_1,\!x_2,\!x_3)\delta(x_2-x_3)\!\Psi^{\dag}(x_1)\Psi^{\dag}(x_2)
\Psi^{\dag}(x_3)\mid \! 0\rangle
\nonumber\\
\mbox{\hspace{-2cm}}&\!\!+&\!\!\!\!\int\limits_0^L\!\!dx_1dx_2dx_3
{\mathrm e}\!^{-{\mathrm
i}\frac{3}{2}\kappa}\chi(x_1,\!x_2,\!x_3)\delta(x_1\!-\!x_2)\Psi^{\dag}(x_1)\Psi^{\dag}(x_2)\Psi^\dag\!(x_3)
\! \mid \! 0\rangle. \label{secondpartb3}
\end{eqnarray}
Now consider the term $b_4$.
Similar to the term $b_3$, we have to move $\Psi(x_3)$ to
the right hand side of $\Psi^{\dag}(x_2)$ in order to form
the state $\Psi^{\dag}(x_1)\Psi^{\dag}(x_2)\Psi^{\dag}(x_3)\mid 0\rangle$.
We thus have
\begin{eqnarray}
\mbox{\hspace{-1.8cm}}
b_4\!\!&=&\!\!\!\!\int\limits_0^L\!\!dx_1dx_2dx_3 {\mathrm
e}\!^{-{\mathrm i}\kappa(\frac{3}{2}+w(x_3,x_1)+
w(x_3,x_2))}\chi(x_1,\!x_2,\!x_3)\!(\Psi^{\dag}\!(x_3))^2\!\!
\left(\Psi^{\dag}\!(x_1)\Psi(x_3){\mathrm e}\!^{-{\mathrm i}\kappa
w(x_3,x_1)}\!\!+\!\delta(x_3\!-\!x_1)
 \right)\!\!
 \Psi^\dag\!(x_2)
  \!\!\mid \!0\rangle \nonumber\\
&=&\!\!\!\!\int\limits_0^L\!\!dx_1dx_2dx_3 {\mathrm e}\!^{-{\mathrm
i}\kappa (\frac{3}{2}+2w(x_3,x_1)+
w(x_3,x_2))}\chi(x_1,\!x_2,\!x_3)\!(\Psi^{\dag}\!(x_3))^2\!\!
\Psi^{\dag}\!(x_1)\Psi(x_3) \Psi^\dag\!(x_2)
  \!\!\mid \!0\rangle \nonumber\\
&+&\!\!\!\!\int\limits_0^L\!\!dx_1dx_2dx_3 {\mathrm e}\!^{-{\mathrm
i}\kappa (\frac{3}{2}+w(x_3,x_1)+
w(x_3,x_2))}\chi(x_1,\!x_2,\!x_3)\delta(x_3\!-\!x_1)
(\Psi^{\dag}\!(x_3))^2
 \Psi^\dag\!(x_2)
  \!\!\mid \!0\rangle \nonumber\\
&\mbox{\hspace{-1.7cm}}=&\mbox{\hspace{-1cm}}\!\!\!\!\int\limits_0^L\!\!dx_1dx_2dx_3
{\mathrm e}\!^{-{\mathrm i}\kappa (\frac{3}{2}+2
w(x_3,x_1)+
w(x_3,x_2))}\chi(x_1,\!x_2,\!x_3)\!(\Psi^{\dag}\!(x_3))^2\!\!
\Psi^{\dag}\!(x_1)\!\! \left(\Psi^\dag\!\!(x_2)\Psi\!(x_3){\mathrm
e}\!^{-{\mathrm i}\kappa
w(x_3,x_2)}\!\!+\!\delta(x_3\!-\!x_2)\!\right)\!
  \!\!\mid \!0\rangle \nonumber\\
&+&\!\!\!\!\int\limits_0^L\!\!dx_1dx_2dx_3 {\mathrm e}\!^{-{\mathrm
i}\kappa (\frac{3}{2}+
w(x_3,x_2))}\chi(x_1,\!x_2,\!x_3)\delta(x_3\!-\!x_1)
\Psi^{\dag}\!(x_1)\Psi^{\dag}\!(x_3)
 \Psi^\dag\!(x_2)
  \!\!\mid \!0\rangle \nonumber\\
&=&\!\!\!\!\int\limits_0^L\!\!dx_1dx_2dx_3 {\mathrm e}\!^{-{\mathrm
i}\kappa (\frac{3}{2}+2 w(x_3,x_1)+
w(x_3,x_2))}\chi(x_1,\!x_2,\!x_3)\delta(x_3\!-\!x_2)(\Psi^{\dag}\!(x_3))^2\!\!
\Psi^{\dag}\!(x_1)\!\mid \!0\rangle \nonumber\\
&+&\!\!\!\!\int\limits_0^L\!\!dx_1dx_2dx_3 {\mathrm e}\!^{-{\mathrm
i}\kappa (\frac{3}{2}+
w(x_3,x_2))}\chi(x_1,\!x_2,\!x_3)\delta(x_3\!-\!x_1)
\Psi^{\dag}\!(x_1)\Psi^{\dag}\!(x_2)
 \Psi^\dag\!(x_3){\mathrm e}^{{\mathrm i}\kappa w(x_3,x_2)}
  \!\!\mid \!0\rangle \nonumber\\
&=&\!\!\!\!\int\limits_0^L\!\!dx_1dx_2dx_3 {\mathrm e}\!^{-{\mathrm
i}\kappa(\frac{3}{2}+2
w(x_3,x_1))}\chi(x_1,\!x_2,\!x_3)\delta(x_3\!-\!x_2)\Psi^{\dag}\!(x_2)\Psi^{\dag}\!(x_3)\!\!
\Psi^{\dag}\!(x_1)\!\mid \!0\rangle \nonumber\\
&+&\!\!\!\!\int\limits_0^L\!\!dx_1dx_2dx_3 {\mathrm e}\!^{-{\mathrm
i}\frac{3}{2}\kappa}\chi(x_1,\!x_2,\!x_3)\delta(x_3\!-\!x_1)
\Psi^{\dag}\!(x_1)\Psi^{\dag}\!(x_2)
 \Psi^\dag\!(x_3)
  \!\!\mid \!0\rangle \nonumber
  \end{eqnarray}
Therefore
  \begin{eqnarray}
b_4&=&\!\!\!\!\int\limits_0^L\!\!dx_1dx_2dx_3 {\mathrm
e}\!^{-{\mathrm
i}\frac{3}{2}\kappa}\chi(x_1,\!x_2,\!x_3)\delta(x_3\!-\!x_2)\Psi^{\dag}\!(x_1)
\Psi^{\dag}\!(x_2)
\Psi^{\dag}\!(x_3)\!\mid \!0\rangle \nonumber\\
&+&\!\!\!\!\int\limits_0^L\!\!dx_1dx_2dx_3 {\mathrm e}\!^{-{\mathrm
i}\frac{3}{2}\kappa}\chi(x_1,\!x_2,\!x_3)\delta(x_3\!-\!x_1)
\Psi^{\dag}\!(x_1)\Psi^{\dag}\!(x_2)
 \Psi^\dag\!(x_3)
  \!\!\mid \!0\rangle. \label{secondpartb4}
\end{eqnarray}

Summing the individual results for $b_1$ (\ref{secondpartb1}), $b_2$
(\ref{secondpartb2}) $b_3$  (\ref{secondpartb3}) and $b_4$
(\ref{secondpartb4}) gives the result
\begin{equation}
B=2\int\limits_0^Ldx_1dx_2dx_3{\mathrm e}^{-{\mathrm
i}\frac{3}{2}\kappa}\chi(x_1,x_2,x_3)(\delta(x_1-x_2)+\delta(x_1-x_3)+\delta(x_2-x_3))
\Psi^{\dag}(x_1)\Psi^{\dag}(x_2)\Psi^{\dag}(x_3)\mid 0 \rangle.
\label{secondpart}
\end{equation}
Finally, we have
\begin{eqnarray}
H\mid \Phi\rangle&=&\frac{\hbar^2}{2m}A+\frac{1}{2}g_{1D}B\nonumber \\
&=&\!\!\int\limits_0^L\!\!dx_1dx_2dx_3 {\mathrm e}\!^{-{\mathrm
i}\frac{3}{2}\kappa}\!\left(\! -\frac{\hbar^2}{2m}
\sum\limits_{i=1}^{3}
\frac{\partial^2\chi}{\partial{x_i}^2}+g_{1D}\!\!\sum\limits_{1\leq
i<j\leq 3}^3\!\!\!\delta(x_i-x_j)\chi
\!\right)\nonumber\\
& &
\times \Psi^{\dag}(x_1)\Psi^{\dag}(x_2)\Psi^{\dag}(x_3)\!\mid 0\rangle \nonumber \\
&=&E\mid \Phi\rangle
\end{eqnarray}
provided $\chi$ satisfies
\begin{equation}
-\frac{\hbar^2}{2m} \sum\limits_{i=1}^{3}
\frac{\partial^2\chi}{\partial x_i^2}+g_{1D}\!\!\sum\limits_{1\leq
i<j\leq 3}^3\!\!\!\delta(x_i-x_j)\chi =E \chi,
\end{equation}
which is the desired three-particle quantum mechanics problem.  
Here $\chi=\chi(x_1,x_2,x_3)$ and $E$ is the eigenvalue.


\section{ Periodic boundary conditions} \label{APP3}


Application of the periodic boundary condition on the Bethe wavefunctions for $N=3$ gives
\begin{eqnarray}
\mbox{\hspace{-1cm}}\chi(x_1=0,x_2,x_3)&=&\mathrm{e}^{-\mathrm{i}\frac{\kappa}{2}(w(x_1,x_2)+w(x_1,x_3)+w(x_2,x_3))}\sum_PA(k_{P1},k_{P2},k_{P3})\mathrm{e}^{\mathrm{i}(k_{P2}x_2+k_{P3}x_3)}\\
\mbox{\hspace{-1cm}}\chi(x_2,x_3,x_1=L)\!\!&=&\mathrm{e}^{-\mathrm{i}\frac{\kappa}{2}(w(x_2,x_3)+w(x_2,x_1)+w(x_3,x_1))}
\!\sum_Q
\!A(k_{Q1},k_{Q2},k_{Q3})\mathrm{e}^{\mathrm{i}(k_{Q1}x_2+k_{Q2}x_3+k_{Q3}L)}
\end{eqnarray}
We thus have
\begin{eqnarray}
e^{\mathrm{i}\frac{\kappa}{2}(2w(x_1,x_2)+2w(x_1,x_3))}
\sum_PA(k_{p1},k_{p2},k_{p3}))\mathrm{e}^{\mathrm{i}(k_{P2}x_2+k_{P3}x_3)}\nonumber\\
=\sum_QA(k_{Q1},k_{Q2},k_{Q3}))e^{\mathrm{i}(k_{Q1}x_2+k_{Q2}x_3+k_{Q3}L)}
\end{eqnarray}
Namely
\begin{eqnarray}
e^{2\mathrm{i}\kappa}\sum_PA(k_{P1},k_{P2},k_{P3}))\mathrm{e}^{k_{P2}x_2+k_{P3}x_3}\nonumber\\
=\sum_QA(k_{Q1},k_{Q2},k_{Q3}))e^{k_{Q1}x_2+k_{Q2}x_3+k_{Q3}L}\label{PBC-1}
\end{eqnarray}
according the order $x_1<x_2\ldots<x_N$.
The condition (\ref{PBC-1}) gives six equations.  
Only three of them
\begin{eqnarray}
\mathrm{e}^{\mathrm{i}k_1L}&=&e^{\mathrm{i}2\kappa}\frac{A(k_1k_2k_3)}{A(k_2k_3k_1)}
=\mathrm{e}^{\mathrm{i}2\kappa}\frac{(k_1-k_2+\mathrm{i}c')(k_1-k_3+\mathrm{i}c')}{(k_1-k_2-\mathrm{i}c')(k_1-k_3-\mathrm{i}c')}\\
\mathrm{e}^{\mathrm{i}k_2L}&=&e^{\mathrm{i}2\kappa}\frac{A(k_2k_1k_3)}{A(k_1k_3k_2)}=\mathrm{e}^{\mathrm{i}2\kappa}\frac{(k_2-k_1+\mathrm{i}c')(k_2-k_3+\mathrm{i}c')}{(k_2-k_1-\mathrm{i}c')(k_2-k_3-\mathrm{i}c')}\\
\mathrm{e}^{\mathrm{i}k_3L}&=&e^{\mathrm{i}2\kappa}\frac{A(k_3k_1k_2)}{A(k_1k_2k_3)}=\mathrm{e}^{\mathrm{i}2\kappa}\frac{(k_3-k_1+\mathrm{i}c')(k_3-k_2+\mathrm{i}c')}{(k_3-k_1-\mathrm{i}c')(k_3-k_2-\mathrm{i}c')}
\end{eqnarray}
are independent.
Similarly consideration of the case
$\chi(x_1,x_2,x_3)=\mathrm{e}^{\mathrm{i}2\kappa}\chi(x_1,x_2+L,x_3)$ gives the same set of equations.

\end{document}